\documentclass[sigconf, 9pt]{acmart}

\def\BibTeX{{\rm B\kern-.05em{\sc i\kern-.025em b}\kern-.08emT\kern-.1667em\lower.7ex\hbox{E}\kern-.125emX}}

\usepackage{xspace}
\usepackage{adjustbox}
\usepackage[ruled]{algorithm2e}
\newcommand{\Kathy}   [1]{{{#1}}}
\newcommand{\Marquita}  [1]{{{#1}}}

\newcommand{\kmer}{{\it k-mer}\xspace}
\newcommand{\kmers}{{\it k-mers}\xspace}
\newcommand{\apriori}{{\it a priori}\xspace}
\newcommand{\lplabel}{{\it Local Processing}\xspace}
\newcommand{\exchlabel}{{\it Exchanging}\xspace}

\newcommand{\ecoli}{{\it E. coli}\xspace}

\begin{document}

%
\title{diBELLA: Distributed Long Read to Long Read Alignment}

%
\author{Marquita Ellis$^{1,2}$, Giulia Guidi$^{1,2}$, Ayd{\i}n Bulu\c{c}$^{1,2}$, Leonid Oliker$^{2}$,  Katherine Yelick$^{1,2}$}
\affiliation{%
\institution{$^1$University of California at Berkeley  \hspace{1in} $^2$Lawrence Berkeley National Laboratory}
}
\email{{mellis, gguidi, abuluc, loliker, yelick}@lbl.gov}

%
\renewcommand{\shortauthors}{Ellis et al.}

%
\begin{abstract}
  We present a parallel algorithm and scalable implementation for
  genome analysis, specifically the problem of finding overlaps and
  alignments for data from ``third generation'' long read sequencers
  \cite{SequencingLandscape}.  While long sequences of DNA 
  offer enormous  advantages for biological analysis and insight,
  current long read sequencing 
  instruments have high error rates
 and therefore require different approaches to analysis than their short
 read counterparts.  Our work focuses on an efficient
 distributed-memory parallelization of an accurate
  single-node algorithm for overlapping and aligning long reads. 
  We achieve scalability of this irregular algorithm by addressing 
  the competing issues of increasing parallelism, minimizing 
  communication, constraining the memory footprint, and ensuring good
  load balance.   The resulting application, diBELLA, is the first distributed
  memory overlapper and aligner specifically designed for
  long reads and parallel scalability.  
  We describe and present analyses for high level design trade-offs and conduct an extensive empirical
  analysis that compares performance characteristics across state-of-the-art HPC systems as well as a commercial cloud architectures, highlighting the advantages of state-of-the-art network technologies. 
\end{abstract}

\copyrightyear{2019} 
\acmYear{2019} 
\setcopyright{usgovmixed}
\acmConference[Authors' preprint for ICPP 2019]{the 48th International Conference on Parallel Processing}{August 5--8, 2019}{Kyoto, Japan} 
\acmPrice{15.00}
\acmDOI{10.1145/3337821.3337919}
\acmISBN{978-1-4503-6295-5/19/08}

%

%
\maketitle
\section{Introduction} \label{sec:intro}

The improved quality, cost, and throughput of DNA sequencing technologies over the past decades has shifted the primary biological challenge from measuring the genome to analyzing the explosion in genomic data, which has far exceeded the growth in computing capabilities.  Yet some of the most complex algorithms for genome analysis are typically run on shared memory machines, limiting parallel scalability, and can run for days or even weeks on large data sets.  Here we present a parallel algorithm and implementation for one such problem involving the latest sequencing technologies and a variety of parallel platforms. 

Because DNA sequencing technologies are unable to read the whole genome in a single run, they return a large amount of short DNA fragments, called {\it reads}.  A read set contains redundant information as each region of the genome is sequenced multiple times (referred to as  {\it depth}, or {\it coverage})  to account for sequencer errors.  
\Kathy{These reads are typically assembled together to form longer genomic regions.
Current sequencing technologies can be divided in two main categories based on the read length: ``short-read'' and ``long-read'' sequencers.
Short-read technologies have very low error rates (well under 1\%)  but the reads are only 100 to 300 base pairs and they cannot resolve repeated regions of the genome longer than those reads \cite{phillippy2008genome, nagarajan2009parametric}.
Long-read technologies, including Pacific Biosciences and Oxford Nanopore, generate reads with an average length over $10,000$ base pairs (bps), but they have error rates from $5\%$ to $35\%$.  

One of the biggest challenges for the analysis of sequencing data is {\it de novo} assembly~\cite{zhang2011practical}, which is the process of eliminating errors and assembling a more complete version of the genome.  This is especially important for plants, animals, and microbial species in which no previously assembled high quality reference genome exists.
The different error rates between short and long reads lead to different approaches to assembly.
For long reads, the first step is typically to find pairs of reads that overlap and resolve their differences (due to errors) by computing the {\it alignments}, i.e., the edits required to make the overlapping regions identical~\cite{nonhybridsurvey, hgap, chin2016phased, nanosense-polish, koren2017canu}.
The read-to-read alignment computation is not limited to genome assembly, and is widely used in various comparisons across or within genomic data sets to identify regions of similarity caused by structural, functional or evolutionary relationships~\cite{mount2004sequence}.
Consequently, highly parallel long-read to long-read alignment would significantly improve the efficiency of these techniques, and enable analysis at unprecedented scale.

In this paper, we focus on this computationally challenging problem of finding overlapping reads and computing their alignment.  We introduce diBELLA, the first long-read parallel distributed-memory overlapper and aligner.  diBELLA uses the methods in BELLA~\cite{guidi2018bella}, an accurate and efficient single node overlapper and aligner that takes advantage of the statistical properties of the underlying data, including error rate and read length to efficiently and accurately compute overlaps.  BELLA is based on a seed-and-extend approach, common to other aligners~\cite{altschul1990basic}, which finds read pairs that are likely to overlap using a near-linear time algorithm and then performing alignments on those pairs.  BELLA parses each read into all fixed-length substrings called \kmers (also called seeds in this context), hashing those \kmers and then finding pairs with at least one common \kmer.  Unlike short read aligners or those that align to a well-established reference, the high error rate in long reads means that BELLA's \kmers must be fairly short (17-mers are typical); this in turn means that some \kmers will appear many times in the underlying genome and can therefore create multiple extraneous overlaps.  diBELLA adopts the innovations from BELLA and parallelizes the seed-and-extend approach by storing \kmers in a distributed hash tables, using that to compute read pairs with a common seed, and then distributing the read pairs for load balanced pairwise alignment.  


diBELLA takes advantage of distributed-memory on high performance computing (HPC) systems as well as commercial cloud environments.
Significant challenges of diBELLA's parallelization include addressing irregular communication, load imbalance, distributed data structures (such as Bloom filters and hash tables), memory utilization, and file I/O overheads.
We demonstrate our scalable solution and detailed performance analysis, across four different parallel architectures, with significantly different architectural design tradeoffs.
In addition, we present communication bounds in terms of input data (genome) and expected characteristics from real data sets.
Our work not only provides a distributed-memory solution for one of the most computationally expensive pieces of the analysis of third-generation sequencing data, it also provides an alternative workload for future architectural developments.  
}

Following some background on the alignment problem in Section~\ref{sec:background}-~\ref{sec:model}, we give a high-level overview of the diBELLA pipeline in Section~\ref{sec:pipeline} and then each of the parallel stages in Sections~\ref{sec:bf}--\ref{sec:alignment}.
In each case, we describe the parallelism opportunities and load balancing challenges with respect to the computation and communication patterns and data volumes. We also show scaling numbers for each stage of the pipeline on the architectures and experimental settings detailed in Section~\ref{sec:measurements}. The architectures include AWS and 3 Cray HPC systems (Edison and Cori at NERSC, and Titan at OLCF). We discuss the overall pipeline performance in Section~\ref{sec:performance}, and conclude with a review of related work in Section~\ref{sec:prior}, and a summary of our conclusions in Section~\ref{sec:conclusions}.

\section{Read-to-Read Alignment} \label{sec:background}

diBELLA computes read-to-read alignment on long-read data to detect overlapping sequences. 
Formally, a {\it pairwise alignment} of sequences {\it s} and {\it t} over an alphabet $\Sigma$ is defined as the pair $(s',t')$ such that $s',t' \in \Sigma \cup \{-\} *$ and the following properties hold:
\begin{enumerate}
  \item  $|s'| = |t'|$, i.e. the lengths of $s'$ and $t'$ are the same
  \item  $\forall_{i=1}^{|s'|}, s_i' \neq -\ OR \ t_i'  \neq -$
  \item  Deleting all ``$-$'' from $s'$ yields $s$, and deleting all ``$-$'' from $t'$ yields $t$.  
\end{enumerate}
Equivalently, we can fix one sequence, $s$, and edit $t$ via insertions and deletions of characters to match $s$.
One is generally interested in only high quality alignments as defined by some scoring scheme that rewards matches and penalizes mismatches, insertions, and deletions.
Finding an {\it optimal} alignment is attainable via a dynamic programming algorithm such as Smith-Waterman and is an $O(|s| \cdot|t|)$ computation~\cite{smith1981identification}.

Pairwise alignment can be extended to sets: given sets of sequences $S$ and $T$, find (all) best pairwise alignments of all $s \in S$ to all $t \in T$.
As a step in {\it de novo} genome assembly, $S$ and $T$ would both correspond to the same set of reads and the pairwise alignment of these two sets would therefore find reads that overlap with each other.
Done naively, set alignment requires $O(|S| \cdot |T| \cdot L^2)$ operations for sequences of length $L$, which becomes intractable for large data sets. However, there are two main improvements possible when performing read-to-read alignment that focus on only high quality outcomes.
\Kathy{First, in place of full dynamic programming for pairwise alignment, one can search only for solutions with a limited number of mismatches (banded Smith-Waterman) and terminate early when the alignment score drops significantly ($x$-drop)~\cite{zhang2000greedy}.  This makes pairwise alignment linear in $L$. diBELLA performs each pairwise alignment on a single node using an $x$-drop implementation from the SeqAn library~\cite{doring2008seqan}.} 
The second improvement involves efficiently finding sequences in $S$ and $T$ that are likely to match before computing the expensive pairwise alignment.
This is accomplished by finding pairs of reads in the input sets that share at least one identical substring.

Each read in $S=T$ is parsed into substrings of fixed length $k$, \kmers, which overlap by $k-1$ characters and are stored in a hash table. 	
\begin{figure}[h]
  \centering
  \includegraphics[width=2in]{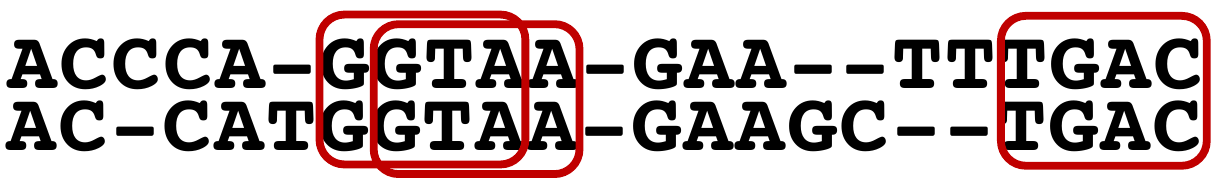}
  \caption{Pairwise alignment of two sequences with $3$ common \kmers of length $4$.} \label{fig:seedexample}
\end{figure}
Figure~\ref{fig:seedexample} illustrates this idea by showing three shared {\it 4-mers} in a given pair of sequences.
Given that long-read data contains errors, the choice of the $k$-mer length is crucial to maximizing the detection of {\it true} overlapping sequences while minimizing the number of attempted pairwise alignments. 	
Quantitatively analyzed in ~\cite{guidi2018bella}, $k$ should be short enough to identify at least one correct shared \kmer between two overlapping sequences, but long enough to minimize the number of repeated \kmers in the genome, which could lead to either spurious alignments or redundant information.
For example, given the two reads in the example in Figure~\ref{fig:seedexample}, a $k$-mer length of $5$ would fail to find an overlap.
Based on the error rate and depth of a given data set, BELLA and diBELLA compute the optimal $k$-mer length to ensure that a pair of overlapping reads will have with high probability at least one correct $k$-mer in common.
A typical \kmer length for long read data sets is $17$-mers based on extensive analysis in~\cite{guidi2018bella}, whereas it is common to use $51$-mers for short read aligners. Note that not all \kmers are useful for detecting overlaps.
\kmers that occur only a single time across $S$ and $T$, called {\it singletons}, are ignored as they are likely erroneous. Even if it wasn't an error, a singleton cannot be used to detect an overlap between two strings since it only occurs in one string.
Conversely, \kmers that occur with very high frequency across the data set are likely from repeated regions of the underlying genome, and can lead to unnecessary or incorrect alignments.
diBELLA therefore eliminates high frequency \kmers over a threshold $m$, which is calculated via the approach presented in BELLA~\cite{guidi2018bella}, using the error rate and other characteristics of the input data set. 
\Kathy{The \kmers that remain after this filtering, we refer to as {\it retained} \kmers and will be used to detect the overlapping reads on which pairwise alignment is performed. This \kmer filtering is specifically for the alignment of long reads to long reads with their high error rates and will affect our parallelization strategy.}

\section{Computational Cost} \label{sec:model}

\begin{figure*}[h]
  \centering
  \includegraphics[width=\textwidth]{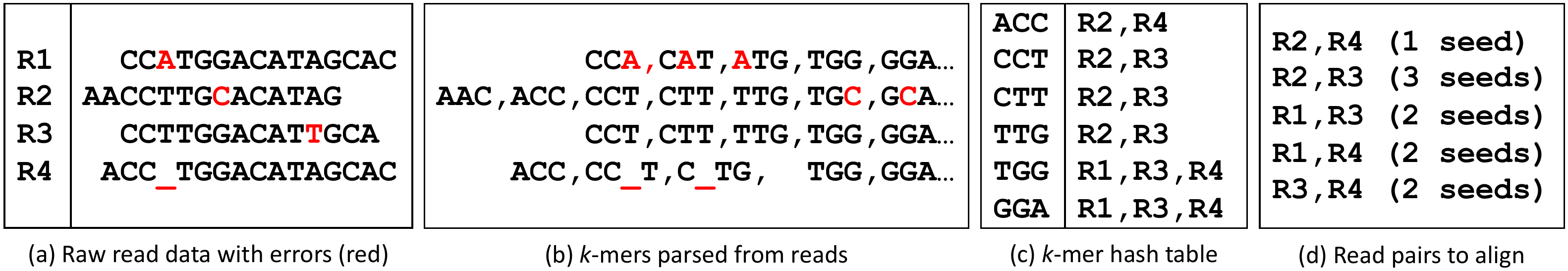}
  \caption{Overview of diBELLA's pipeline, using $k = 3$ as example: (a) raw input data, (b) $k$-mer extraction, (c) $k$-mer hash table and associated read list, and (d) read pair alignment using the seed-and-extend paradigm.} \label{fig:pipelineexample}
\end{figure*}

To approximate the computational cost, we first note that the size of the long read input data set $N$ from a given genome is determined by two variables, the size of the underlying genome $G$ and the average depth of per base coverage $d$ (equation~\ref{eq:input_size}). 
\begin{equation}\label{eq:input_size}
N = G \cdot d
\end{equation}
If $L$ is the average length of sequences in the input, then the size of the read set is $R=G \cdot d/L$.
The computation extracts \kmers starting at every location in each read of the input set.
Thus a read of length $L$ has $L-k+1$ \kmers (although not necessary unique ones).
For long-read data, $L$ is generally in the range $1,000-100,000$ and $k$ is in the range $11-21$ so we approximate the number of \kmer's per read as $L$.
The number of the \kmers parsed from the input (i.e., the {\it bag} of \kmers, which may have duplicates) is thus approximately $G \cdot d$.
\begin{equation}\label{eq:kmer_cardinality}
\frac{G \cdot d \cdot (L-k+1)}{L}\approx G \cdot d
\end{equation}

Therefore, the total volume of \kmers from the input is $k\cdot G \cdot d$ characters.
Each \kmer character from the four letter alphabet $\{A,C,T,G\}$ can be represented with $2$ bits.
To support varying values of $k$ with efficient memory storage and alignment, we provide compile time parameters for the \kmer representation (typically set to $32$ bits or the nearest larger power of two).
In general, we avoid storing the entire \kmer bag in memory at once, unless sufficient distributed memory resources happen to be available. For example, the \kmer bag size of two PacBio \ecoli data sets, with 30x and 100x coverage respectively, is over 2 billion and 7 billion characters. 

diBELLA operates predominately on the much smaller set of distinct \kmers, although we retain some information about each instance of a \kmer, such as its locations within different reads.
We further reduce the size of the retained \kmer set (as described in Section~\ref{sec:background}) by eliminating singletons and high-frequency \kmers. 
\Kathy{Assuming a properly chosen value of $k$, the set of filtered \kmers is approximately the size of the final assembled genome $G$.}

\section{\lowercase{di}BELLA Overview}\label{sec:pipeline}
Our distributed-memory diBELLA design is a multi-stage parallel pipeline. \kmers are first extracted from files of reads and filtered by frequency, as described in Section~\ref{sec:background}. Each processor manages a subset of the reads and a subset of the \kmers. Note that there is no inherent locality in the order of the reads from the input files (called  FASTQ) with respect to their overlap. The first phase builds a distributed Bloom filter\cite{egeorganas2016} to identify and eliminate most singleton \kmers. The second phase builds a hash table of non-singleton \kmers (as approximated by the Bloom filter), and further filters \kmers exceeding the high occurrence threshold, $m$ (see Section \ref{sec:background}). The remaining hash table represents a graph with reads (represented by identifiers) as vertices and reliable \kmers as edges. That is, two vertices (long reads) are connected if they share a common \kmer that was retained after filtering. The next stage forms all pairs of read IDs that share a retained \kmer and tracks their location within the reads. \Kathy{The final stage performs alignment on these read pairs using the shared \kmer as the starting position (seed) for pairwise alignment.}

Our distributed memory design is a four-stage pipeline, with an example shown in Figure~\ref{fig:pipelineexample} : 
\begin{enumerate}
\item Extract \kmers from files of reads and store in a distributed Bloom filter to eliminate singleton \kmers. Initialize the hash table with non-filtered \kmers.
\item Extract \kmers and their location metadata from the files again. Insert into the distributed hash table only if the \kmer is already resident.  After this is done, remove singleton \kmers that were missed by the Bloom filter and those that exceed the high occurrence threshold, $m$. 
\item For each \kmer in the hash table, take the associated list of read IDs (and positions) and form all pairs of reads, \Kathy{assigning each pair to one processor}.
\item \Kathy{Redistribute and replicate reads (the original strings) to match read-pair distribution and perform pairwise alignment on each pair locally.}
\end{enumerate}

The algorithm makes two passes over the data in order to not store all the parsed \kmers in main memory; diBELLA executes in a streaming fashion \Kathy{with a subset of input data at a time} to limit the memory consumption. 


The Bloom filter, hash table, and list of read pairs are all distributed across the nodes, and the predominate communication pattern, common to each stage, is irregular all-to-all exchanges. The first two stages exchange \kmers for counting and for initializing the hash table with \kmers and respective source locations. \Marquita{The \kmers are mapped to processors uniformly at random via hashing, such that each processor will own roughly the same number of distinct \kmers, as in \cite{egeorganas2016}. Further details for these stages follow in Sections~\ref{sec:bf}-\ref{sec:ht}.}
The third phase consolidates read-pairings ({\it overlaps}), and their lists of shared \kmer positions\Marquita{, which represent alignment tasks. The details of the parallelization and task redistribution are provided in Section~\ref{sec:overlap}.
The final stage computes all pairwise alignments.} Because the pairwise alignments require the full reads, any non-local reads are requested and received by the respective processor. \Marquita{This last stage is described in detail in Section~\ref{sec:alignment}.}
Overall, our design employs Bulk Synchronous Parallelism \cite{ValiantBSP1990} throughout, with the communication implemented via MPI Alltoall and Alltoallv functions. Note that a load imbalance can result from the data characteristics, including highly repetitive genome regions. \Kathy{The current diBELLA implementation makes particular design choices for data layout, communication aggregation, and synchronization, and we evaluate their effectiveness through extensive cross-platform performance analysis while identifying opportunities for future optimizations within the general framework. The specific techniques for \kmer length selection, filtering and local alignment are based on those in BELLA, but the parallelization approach is applicable to this general style of long read aligner based on \kmer filtering and hashing.} 

\section{Experimental Setup} \label{sec:measurements}


\Kathy{Our experiments were conducted on four computing platforms, which include  
HPC systems with varying balance points between communication and computation, as well as a commodity AWS cluster.  This gives us performance insights into tradeoffs between extremes of network capabilities. 
Evaluated platforms include the Cori Cray XC40 and Edison Cray XC30 supercomputers at NERSC, the Cray XK7 MPP at the Oak Ridge National Lab, and an Amazon Web Services (AWS) c3.8xlarge cluster.
Details about each architecture are presented in Table~\ref{tab:machinetab}.  
Titan has GPUs and CPUs on each node, but we use only the CPUs with total 16 (integer) cores per node.
AWS does not reveal specifics about the underlying node architecture or interconnect topology,
other than an expected 10 Gigabit injection bandwidth. 
Based on our measurements, the AWS node has similar performance to a Titan CPU node. 
Both data sets are small enough to fit in the memory of a single node, and in all experiments, MPI Ranks are pinned to cores.
}
\begin{table*}[t]
\centering
\caption{Evaluated platforms. $^\ast$128 byte Get message latency in microseconds. $^\dagger$Using
the optimal number of cores per node. $^\ddagger$Measured over approx. 2K cores or maximum
(128 for ethernet cluster). $^\S$MB/s with 8K message sizes. $^\alpha$CPU nodes only.
} \label{tab:machinetab}

\begin{tabular}{llll}
\hline\noalign{\smallskip}
Processor & Cori I Cray XC40 & Edison Cray XC30 & Titan Cray XK7$^\alpha$ \\ 
		  & Intel Xeon (Haswell) & Intel Xeon (Ivy Bridge) & AMD Opteron 16-Core \\ 
\hline\noalign{\smallskip} 
Freq (GHz)						& 2.3	& 2.4	& 2.2		\\
Cores/Node						& 32	& 24	& 16		\\
Intranode LAT$^{\ast\dagger}$	& 2.7	& 0.8	& 1.1		\\
BW/Node$^{\dagger\ddagger\S}$	& 113.0	& 436.2	& 99.2		\\
Memory (GB)						& 128	& 64	& 32		\\
Network and Topology			& Aries Dragonfly	& Aries Dragonfly	& Gemini 3D Torus		\\
\hline
\end{tabular}
\end{table*}

To highlight cross-network performance and communication bottlenecks, most of our experiments use an input data set and runtime parameters that result in low computational intensity. This data
is from \ecoli bacteria with a depth of $30\times$, which consists of $16,890$ long reads from the from {\it Escherichia coli} MG1655 strain, resulting in a $266$ MB input file; it has been sequenced using PacBio RS II P5-C3 technology and it has an average read length of $9,958$ bp.  
The second data set, \ecoli $100\times$, was sequenced using PacBio RS II P4-C2 and uses a depth of $100$. It consists of $91,394$ long reads from the same strain with an average read length of $6,934$ bps, resulting in a $929$ MB input file.
diBELLA's overlap detection step identifies $2.27$M potentially overlapping read pairs for the first data set and $24.87$M for the second one.

\Kathy{Computational intensity is most affected by the number of alignments performed for each pair of reads, since each pair might share varying numbers of seeds.  Some of these seeds reflect a shifted version the same overlapping region, whereas others may be  independent (and ultimately incorrect) overlaps.}
We use three different options to provide a range of computational intensity.  At the two extremes, the {\it one-seed} option computes pairwise alignment on exactly one seed per pair,
while the {\it all-seed} option computes pairwise alignment on all the available seeds separated by at least the \kmer length.  As an intermediate point we consider only seeds separated by $1,000$ bps. 
\Marquita{The analysis associated with the design of BELLA~\cite{guidi2018bella}, shows that even 1,000 bp separation can be used without significantly impacting quality.}

Both data sets are sufficiently small that the working set size fits on a single node across the platforms in our comparison. This choice enables us to show the performance impact of intra-node to inter-node communication on the overall pipeline performance and highlight scaling bottlenecks, and to explore strong scaling on a modest number of nodes, important for comparison with AWS.
\section{Bloom Filter Construction} \label{sec:bf} 
Given that singleton \kmers constitute the majority of the \kmer data set, retaining them is memory inefficient since it would require storage $k$ times larger than the input size. Therefore the goal of this stage is to build a distributed Bloom filter to identify (with high probability) singleton \kmers, which can be ignored. It also enables the initialization of a distributed hash table containing the unfiltered \kmers. Briefly, a Bloom filter is an array of bits that uses multiple hash functions on each element to set bits in the array.  Due to collisions, a value may not be in the array even if its hash bits are set, but a value with at least one zero is guaranteed to be absent from the set~\cite{bloom1970space} (i.e. it may allow false positives, but does not contain false negatives).  We follow the methodology of the HipMer short read assembler~\cite{georganas2018hipmer} for this stage, but note that the Bloom filter is even more effective for long reads due to their higher error rate --- up to $98\%$ of \kmers from long reads are singletons vs. $60-85\%$ for short reads. 
Minimizing the Bloom filter false positive rate depends on the (unknown \apriori) cardinality of the \kmer set. In our experiments thus far, we have not encountered a case where approximating the \kmer cardinality using equation~\ref{eq:kmer_cardinality} and typical ratios of singleton \kmers to all \kmers across data sets did not provide a sufficiently accurate estimate, such that the more expensive HyperLogLog algorithm in HipMer was required. However, we suspect that for extremely large (tens of trillions of base pairs) and repetitive genomes that we may encounter the same issues that led to this optimization in HipMer.

As mentioned, the input reads are distributed roughly uniformly over the processors using parallel I/O, but there are is no locality inherent in the input files. 
Each rank in parallel parses its reads into \kmers, hashes the \kmers, and eventually sends them to a processor indicated by the hash function. \Marquita{The hash function ensures that each rank is assigned roughly the same number of \kmers.} On the remote node, the received \kmers are inserted into the local Bloom filter partition. If a \kmer was already present, it is also inserted into the local hash table partition. Although all $G \cdot d$ \kmers are to be computed, this process is performed in stages since only a subset of \kmers may fit in memory at one time. 
The Bloom filter construction communicates nearly all (roughly $(P-1)/P$) of the \kmer instances to other processors \Marquita{in a series of bulk synchronous phases. The total number of phases depends on the size of the input, and the irregular all-to-all exchange is implemented with MPI Alltoall and Alltoallv.} After the hash table is initialized with \kmer keys, the Bloom filter is freed.


%

\begin{figure}[!tb]
  \centering 
  \includegraphics[width=\linewidth]{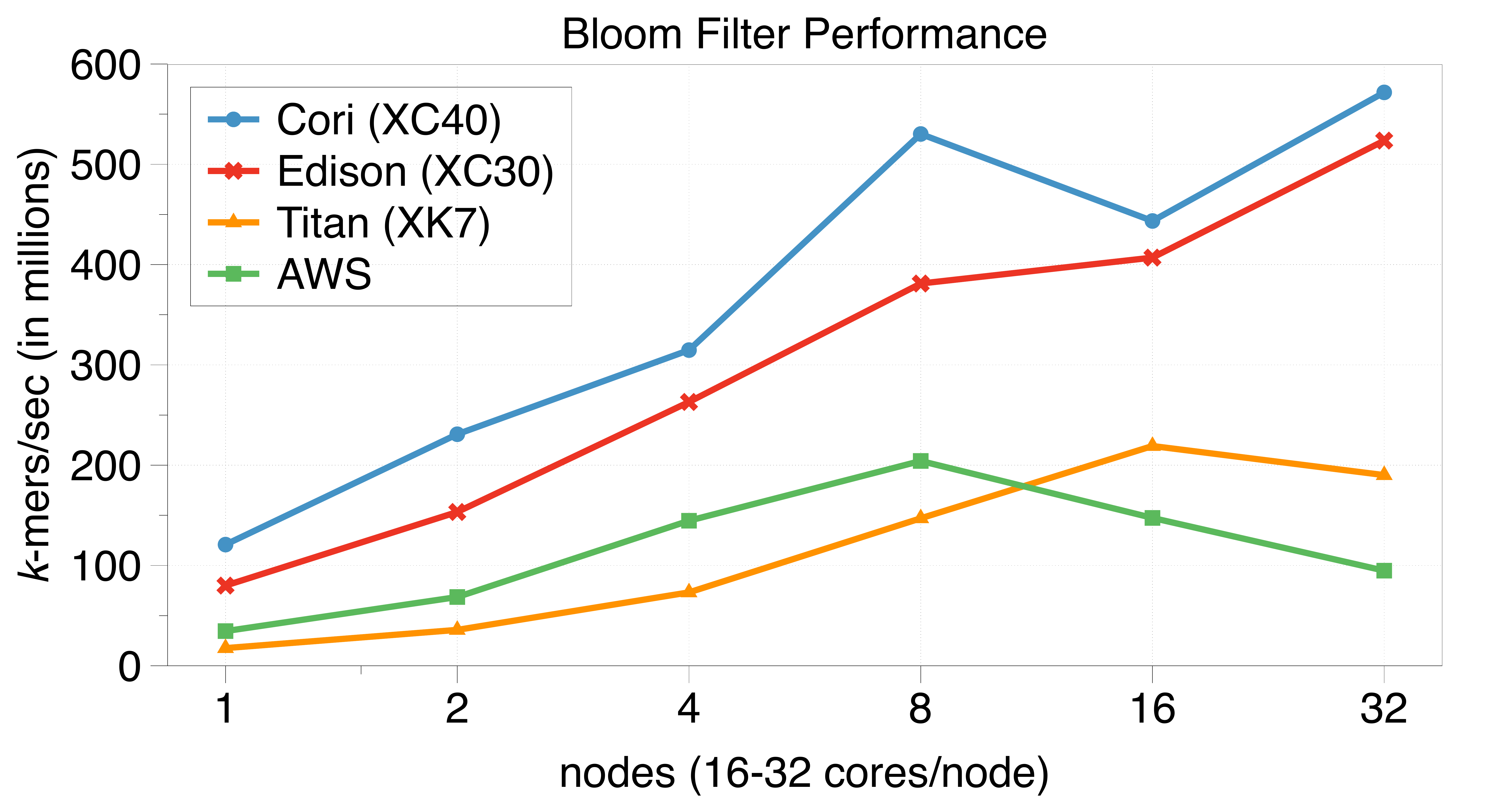}
  \caption{Bloom Filter cross-architecture performance in millions of \kmers processed / sec, given E.coli 30x one-seed.}%
 \label{fig:bf_rate}
\end{figure}

Figure~\ref{fig:bf_rate}, shows strong scaling performance (including communication) of the Bloom filter phase across the four platforms in our study, measured in millions of \kmers processed per second. Note that on Titan (Cray XK7), 1 MPI Rank is assigned to each {\it Integer Core}/L1 Cache, and the GPU's are not utilized. Each node of Titan contains 16 {\it Integer Cores}, the overall computational peak of which is significantly lower than Cori and Edison (which contain 32 and 24 more powerful cores per node, respectively). Titan's \kmer processing rate is most similar to the AWS cluster (which contains 16 cores per node), and surpasses AWS performance only when communication becomes the dominant bottleneck at 16-32 nodes.

\begin{figure}[!tb]
  \centering 
  \includegraphics[width=\linewidth]{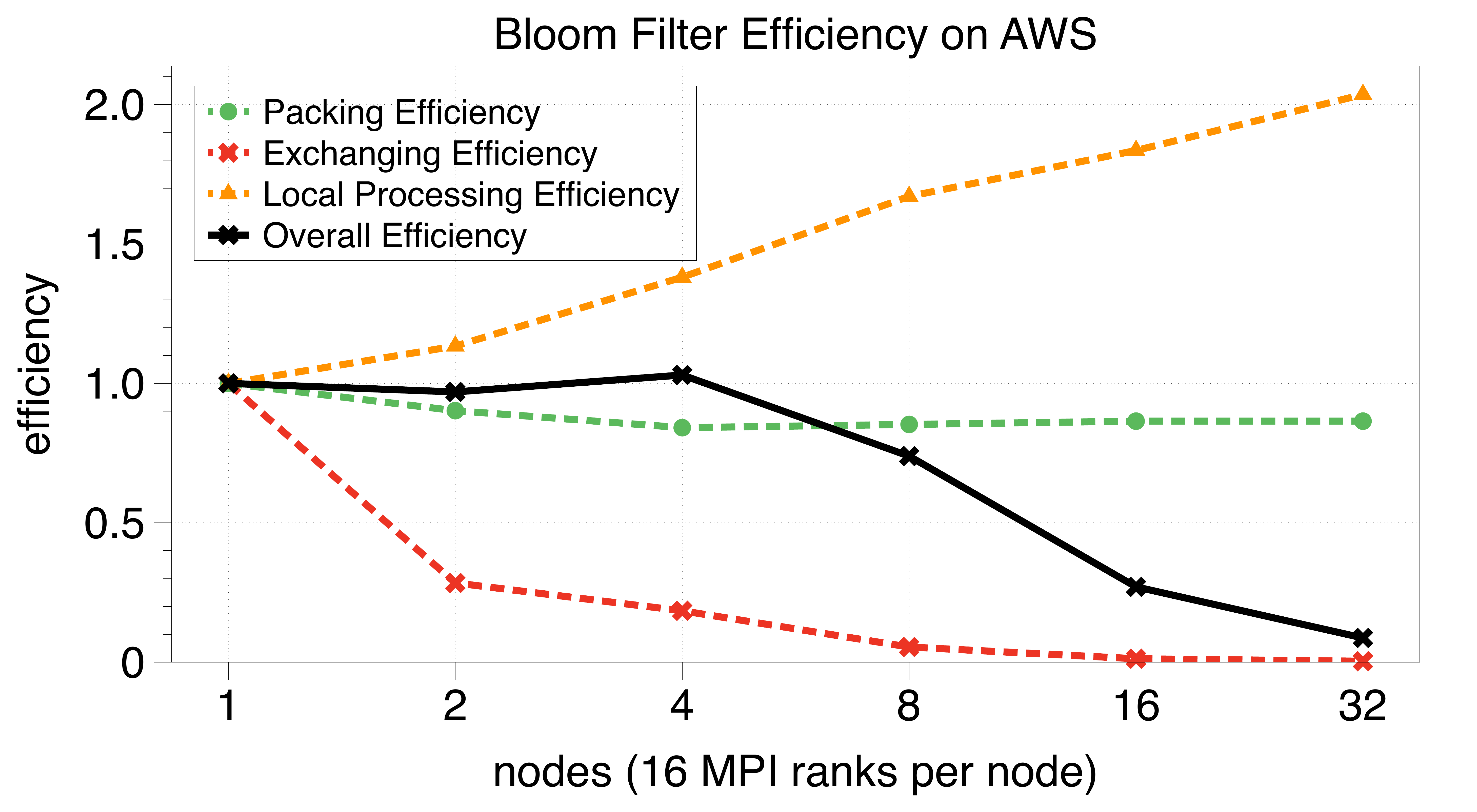}
  \caption{Bloom Filter efficiency on AWS within a 32 node placement group, 1 MPI Rank per core, 16 cores per node, strong scaling with E.coli 30x one-seed.}%
  \label{fig:bf_efficiency_AWS}
\end{figure}

Figure~\ref{fig:bf_efficiency_AWS} presents a detailed breakdown of the strong scaling efficiency on AWS. Note that the \lplabel (hashing and storing \kmers) speeds up  superlinearly, since more of the input fits in cache for this strong scaling experiment. On the other hand, the \exchlabel efficiency, computed relative to the single (intra) node communication, degrades significantly with increased concurrency, and eventually overwhelms the overall runtime. More detailed measurements (not shown) reveal that some of the poor scaling in \exchlabel is only in the first call to MPI's Alltoallv routine. The overhead is assumably from the MPI implementation's internal data structure initialization, related to process coordination and communication buffers setup for subsequent calls. 


\section{Hash Table Construction}\label{sec:ht}

\begin{figure}
  \centering 
  \includegraphics[width=\linewidth]{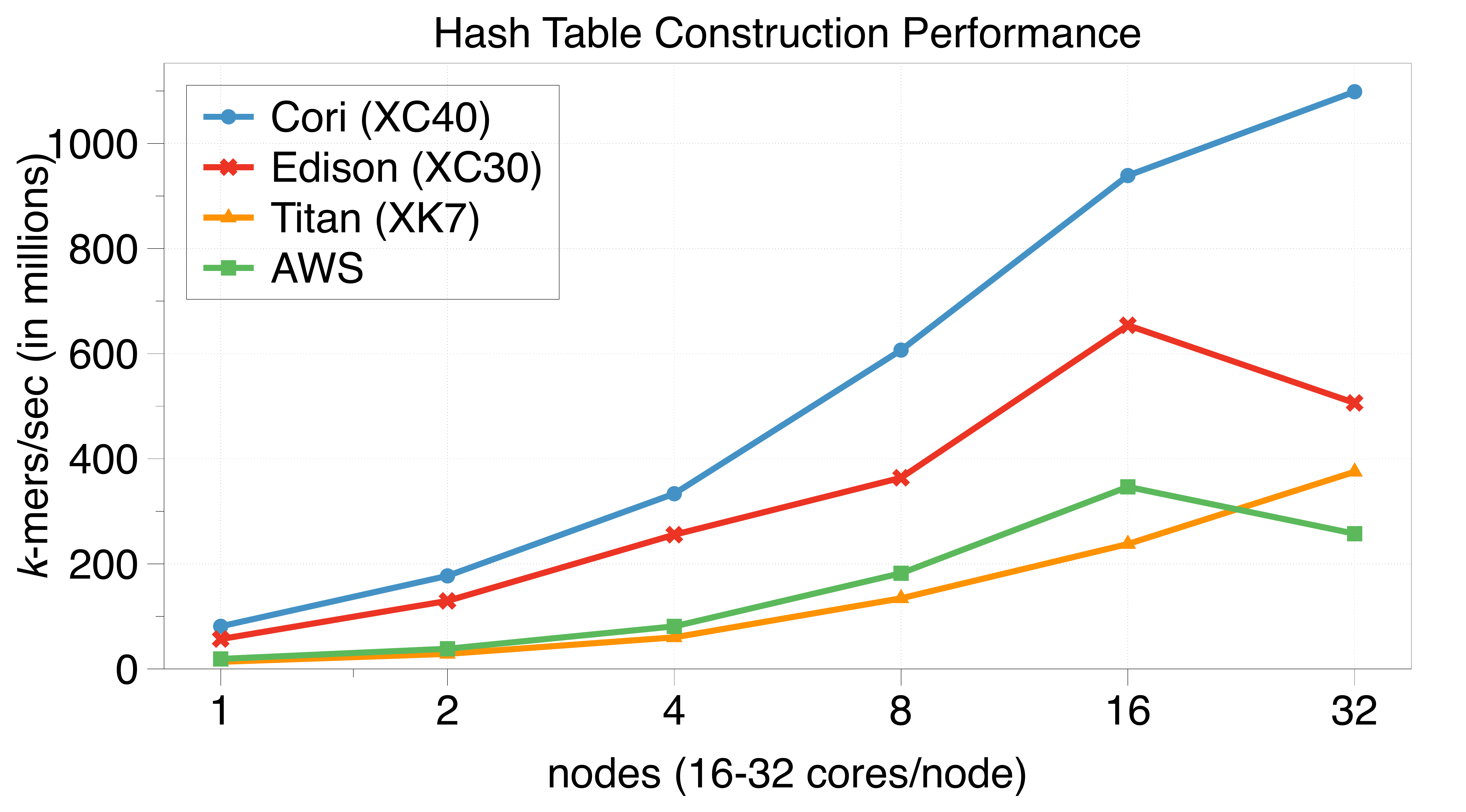}
  \caption{Hash Table stage, cross-architecture performance in millions of \kmers /second given E.coli 30x one-seed.}%
  \label{fig:ht_rate}
\end{figure}

In order to identify reads with at least one common \kmer, the next phase builds a hash table of \kmers and the lists of all read identifiers (RID's) and locations at which they appeared. 
 In this stage all reads are again parsed into \kmers, hashed, and sent to the processor owning that \kmer, and if the \kmer key exists in the hash table (not a singleton), it is inserted with its RID and location and its count is incremented. \Marquita{The same strategy for load balancing \kmers as in the Bloom filter construction stage (Section~\ref{sec:bf}) is employed here; the \kmers are hashed to the same distributed memory locations that they were in the previous stage.} At the end of this process, the local hash table partitions are traversed independently in parallel to remove any \kmers that occur more times than the maximum frequency, and any false-positive singletons. The remaining \kmers are referred to as  {\it retained} \kmers. This extra RID and location information makes the hash table different than other tools intended for de Bruijn graph construction of short reads (such as HipMer~\cite{georganas2018hipmer}) or those used to analyze read data directly by counting \kmers (such as Jellyfish~\cite{marccais2011fast}). The communication is again done in a memory-limited set of bulk-synchronous phases, \Marquita{where the irregular all-to-all exchange of \kmers and associated data is implemented with MPI Alltoall and Alltoallv}. 
Note that while the communication volume of this stage is 2.5x larger than the Bloom filter stage, the amount of computation is also higher due to the RID and location handling, as well as the hash table traversal. This difference is apparent in the strong scaling performance comparison between Bloom filtering in  Figure~\ref{fig:bf_rate} and hash table construction in Figure~\ref{fig:ht_rate}. 
Although the trends are similar across stages and platforms, the computation rate of the hash table stage is roughly double that of the Bloom filter stage. Once again improved cache behavior results in superlinear speed up for this strong scaling computation.

\section{Overlap}\label{sec:overlap}
Once the distributed hash table is computed, which maps reliable \kmers to source locations (RIDs and positions), the overlap computation is straightforward. \Marquita{Rather than constructing the matrix explicitly as in BELLA, we avoid the associated overhead and compute overlaps directly from hash table partitions, independently in parallel in diBELLA. Further exploration of the associated design tradeoffs is part of ongoing work. }
Algorithm~\ref{alg:overlap_loop} illustrates this simple, direct computation of the set of all pairs of reads represented by identifiers $(r_a, r_b)$, where $r_a$ and $r_b$ share reliable \kmer(s). Each \kmer ``contributes'' to the discovery of [2, $m (m-1)/2$] read pairs where $m$ is the maximum frequency of {\it reliable} \kmers in diBELLA, or simply the maximum frequency of retained \kmers in general.  \Marquita{ Each of these represents an alignment task for the next stage. However, the owner of the \kmer matching $(r_a, r_b)$ may not be the owner of either involved read. To maximize locality in the alignment stage (minimize the movement of reads) each task is buffered for the owner of $r_a$ or $r_b$ (which may be the same owner), according to the simple odd-even heuristic in Algorithm~\ref{alg:overlap_loop}. 
Recall, reads in the input are unordered and partitioned uniformly. The hash table values (RID lists) are also unordered. Hence, for fairly uniform distributions of reliable \kmers in the input, we expect this heuristic to roughly balance the number of alignment tasks assigned to each processor. }
Load balancing by number of tasks is however imperfect, since individual pairwise alignment tasks may have different costs in the alignment stage.
The computational impact of various features, such as read lengths and \kmer similarity, could be used for estimating the cost changes within the pairwise alignment kernel. We leave further analysis of the relationship between the choice of pairwise alignment kernel and overall load balancing to future work. \Marquita{Our expectations of the general load balancing strategy are discussed further with empirical results in the context of the alignment stage description, Section~\ref{sec:alignment}. The final steps of the overlap stage are the irregular all-to-all communication of buffered tasks, implemented with MPI\_Alltoallv, and the (optional) output of the overlaps. }

\begin{algorithm}
\SetAlgoLined
\caption{Parallel (SPMD) hash table traversal } \label{alg:overlap_loop}
\KwResult{All pairs of reads sharing at least 1 retained \kmer in hash table partition, H, \Marquita{and corresponding \kmer positions (elided) are composed into alignment tasks. Each task, with read identifiers $(r_a,r_b)$, is stored in a message buffer for the owner of $r_a$ or $r_b$}.   
}
\For{each \kmer key $k_{hash}$ in hash table \bf{H}}{
  \For{i = 0 to m-2}{
    \For{j=i+1 to m-1}{
    	$(r_a,r_b)$ = task(H[$k_{hash}$][i] , H[$k_{hash}$][j],...)\\
	\uIf{ $r_a \% 2 = 0$ AND $r_a > r_b+1$ }{ 
	  buffer[owner($r_a$)]$\gets (r_a,r_b) $
	}
	\uElseIf{$r_a \% 2 \neq 0$ AND $r_a < r_b+1$}{
	  buffer[owner($r_a$)]$\gets (r_a,r_b) $
	}
	\Else{buffer[owner($r_b$)]$\gets (r_a,r_b) $}
    }
  }
}
\end{algorithm}

Neither the number of overlapping read pairs nor the number of retained \kmers common to each can be determined for a given workload until runtime. However, we provide generalizable bounds on the computation and communication from a few basic observations.
Recall from Section \ref{sec:background} that the total number of \kmers parsed from the input is one order of magnitude larger than the number of characters in the input.
However, only a small fraction of these are stored, those that are distinct, and with frequency in [2, $m$]. 
Let the fraction of retained \kmers to the total number of (input) \kmers ($K_{input}$) be $\iota_{input}$, and the fraction of retained \kmers to the size of the \kmer set, $|K_{set}|$ be $\iota_{set}$. The importance of the distinction is that $|K_{set}|$ cannot be known until the end of the \kmer analysis stage, whereas $K_{input}$ is known \apriori. Further, $|K_{set}| \le K_{input}$, and thus $\iota_{set} \ge \iota_{input}$. In our cross-genome analysis, $\iota_{set} \in [0.04,0.12]$. 
This analysis is useful for estimating the overlap computation and communication costs, and applicable beyond our particular implementation.

An upper bound on the total (global) number of overlaps follows in Equation~\eqref{eq:num_overlaps}. 
The lower bound (Equation~\eqref{eq:num_overlaps_lower}) follows from the fact that retained \kmers must occur in at least two distinct reads (identifying at least one overlap) or they are discarded.
The parallel computational complexity of Algorithm~\ref{alg:overlap_loop}'s overlap detection (with $P$ parallel processors) is shown in Equation~\eqref{eq:overlap_comp}, which assumes constant-time storage of read pair identifiers. The hidden constant in Equation~\eqref{eq:overlap_comp} is halved by exploiting asymmetry.

\begin{equation}\label{eq:num_overlaps}
O(\iota_{set} \times K_{set} \times m^2 ) < O(\iota_{input} \times K_{input} \times m^2)
\end{equation}
\begin{equation}\label{eq:num_overlaps_lower}
O(\iota_{set} \times K_{set} ) < O(\iota_{input} \times K_{input})
\end{equation}
\begin{equation}\label{eq:overlap_comp}
O(\frac{\iota_{set} K_{set} m^2}{P})
\end{equation}

Ignoring the constant for the size of the overlap representation (a pair of read identifiers and positions in our case), the aggregate communication volume is also bounded above by Equation~\ref{eq:num_overlaps}, and below by Equation~\ref{eq:num_overlaps_lower}. 

As a last computational step, after the overlaps are computed and communicated (and lists of common \kmers consolidated), the lists may be filtered further depending on certain runtime parameters. That is, some subset of all \kmers per overlapping read pair will be used to seed the alignment in the next stage; the subset is determined by the shared retained \kmers total (simply all may be used) and also by certain runtime parameters which can be thought of as ``exploration'' constraints. These include the minimum distance between seeds, and the maximum number of seeds to explore per overlap. A discussion of these settings in relation to alignment accuracy versus computational cost is presented in the BELLA analysis\cite{guidi2018bella}. In general, increasing the number of seeds to explore per overlap increases computational cost of the alignment stage (not necessarily linearly), depending on the pairwise alignment kernel employed. We present results varying the number of seeds Section \ref{sec:alignment}.

\begin{figure}
  \centering
  \includegraphics[width=\linewidth]{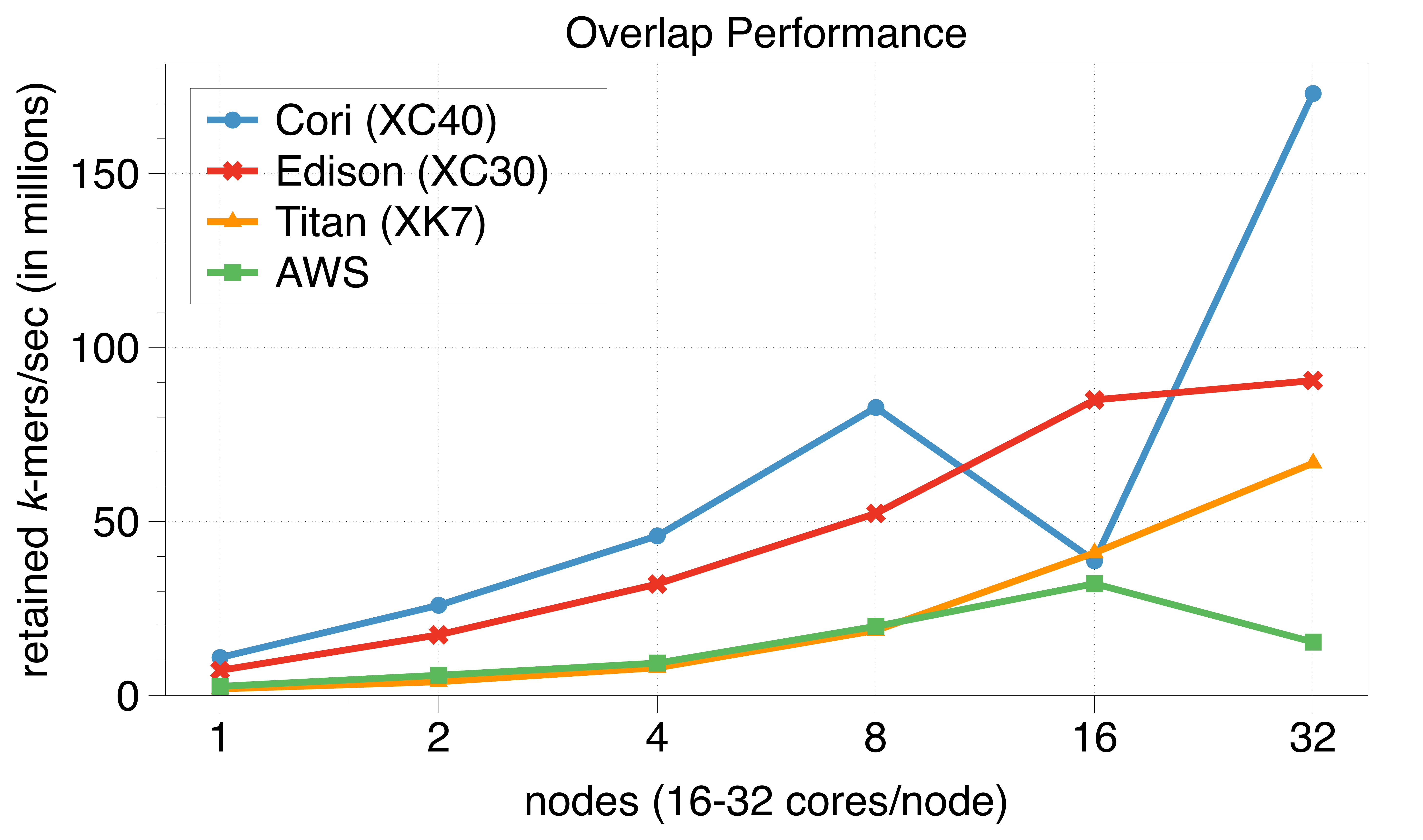}
  \caption{Cross-architecture Overlap stage performance in millions of retained \kmers /second given E.coli 30x one-seed.}%
  \label{fig:overlap_rate}
\end{figure}

Strong scaling results for the overlap stage are shown in Figure \ref{fig:overlap_rate}. These are presented across our evaluated platforms in terms of millions of retained $k$-mers processed per second, and show similar computational behavior as the previous stages. One unexpected feature of this graph is the dip Cori's performance trend at 16 nodes, due to an unexpected spike in the communication exchange time that does not continue to 32 nodes. The absolute time is short enough to have been caused by interference in the network, but the spike nonetheless brings the Cori performance down to Titan and AWS's at 16 nodes.  

\section{Alignment}\label{sec:alignment}
The \kmer load balancing strategy described in sections \ref{sec:bf}-\ref{sec:ht} enables uniform \kmer load balancing and complete parallel overlap detection. The rebalancing of overlapping reads, however, is left for the alignment stage. Recall, that the input reads are not ordered, and our algorithm partitions them as uniformly as possible at the beginning of the computation (by the read size in memory). Only \kmers and read identifiers (not the actual reads) are communicated in the initial stages of the pipeline.  After the overlap stage communication, each overlap identifier together with the associated list of share \kmer positions, are stored in the appropriate owners location.  Note, the \kmer positions are retained rather than recomputed because they are the locations of (globally) rare \kmers (see Section~\ref{sec:background}). Computing the alignment of any overlapping pair of reads, however, requires both of the respective input reads. 

The properties of the overlap graph underpin the communication design of our application. The size of the retained \kmer set determines the size (and sparsity) of the overlap graph. From our filtering steps, we expect this graph to be sparse; from empirical observations across data sets, the filtering typically reduces the \kmer set size by 85-98\%. 

To effectively maximize locality and bandwidth utilization under these conditions, we first explore the performance of a bulk synchronous exchange implemented via MPI\_Alltoallv. Note that once the reads are communicated, the alignment computation can proceed independently in parallel.
We expect that speedups from the subsequently embarrassingly parallel alignment computations (which are quadratic for exact pairwise alignment and at least linear in the length of the long reads 
for approximate alignments) will compensate for inefficiencies in the communication to some  workload-dependent degree of parallelism. 

\begin{figure}
  \centering
  \includegraphics[width=\linewidth]{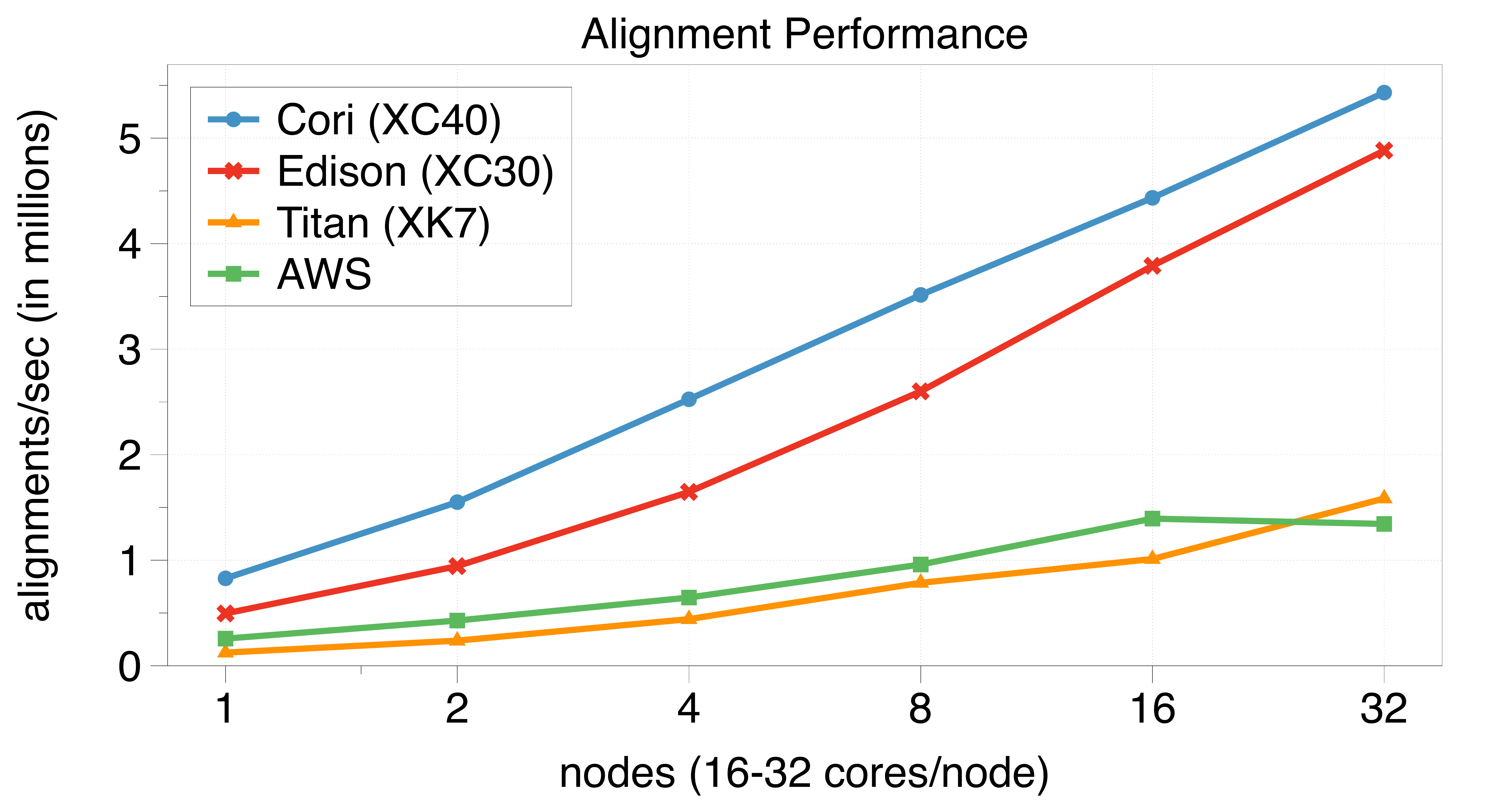}
  \caption{Cross-architecture Alignment stage strong scaling in millions of alignments / second given E.coli 30x one-seed. }%
  \label{fig:AlignmentRateAll}
\end{figure}

Each overlap shares an unknown \apriori number of retained \kmers that are used as alignment seeds. Anywhere between one or all of these seeds will be explored in application runs, depending on the user's objectives and runtime settings. Figure \ref{fig:AlignmentRateAll}  shows performance (alignments per second) across our evaluated platforms machines using the (computationally worst-case) one seed per alignment. Here, the number and speed of the cores per node determine the relative performance ranking (see Table~\ref{tab:machinetab}), with Cori's 32 cores/node clearly surpassing the other systems. 

\begin{figure}
  \centering
  \includegraphics[width=\linewidth]{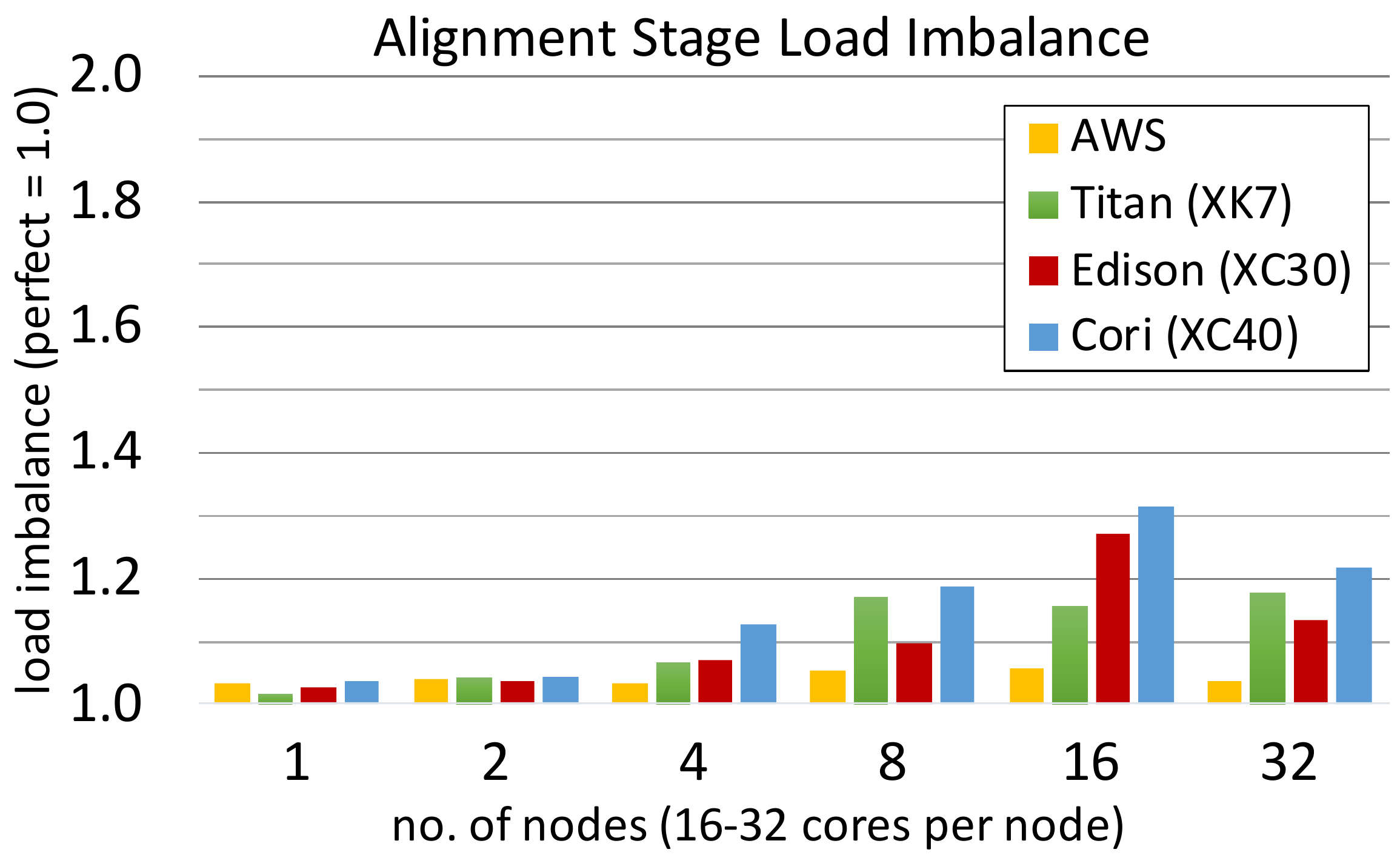}
  \caption{Alignment stage load imbalance on E.coli 30x one-seed, calculated using maximum over average stage times across ranks (1.0 is perfect).}%
  \label{fig:AlignmentLoadImbalance}
\end{figure}

The load balancing strategy described in Sections~\ref{sec:overlap}-\ref{sec:alignment} produces near perfect load balancing in terms of the number of alignments computed per parallel process, but imperfect load balancing in terms of time to exchange and compute all alignments. Figure \ref{fig:AlignmentLoadImbalance} shows the latter load imbalance, calculated as maximum per rank alignment stage times over average times across ranks (1.0 is perfect). There are two reasons for this load imbalance in terms of compute and exchange costs: (1) reads have different lengths, which effect both the exchange time and the pairwise alignment time, (2) the x-drop algorithm returns much faster when the two sequences are divergent because it does not compute the same number of cell updates. A smarter read-to-processor assignment could optimize for variable read lengths, eliminating the exchange imbalance. However, the imbalance due to x-drop can not be optimized statically as it is not known before the alignment is performed. To mitigate the impact of (2), one would need dynamic load balancing, which is known to be high-overhead in distributed memory architectures. The load imbalance in terms of the number of alignments performed per processor is less than $0.002\%$ across all machines and scales. Future work should consider not only the number of alignments per processor but other kernel-dependent characteristics affecting the cost of each pairwise alignment.


\section{Performance Analysis}\label{sec:performance}

\begin{figure}
  \centering 
  \includegraphics[width=\columnwidth]{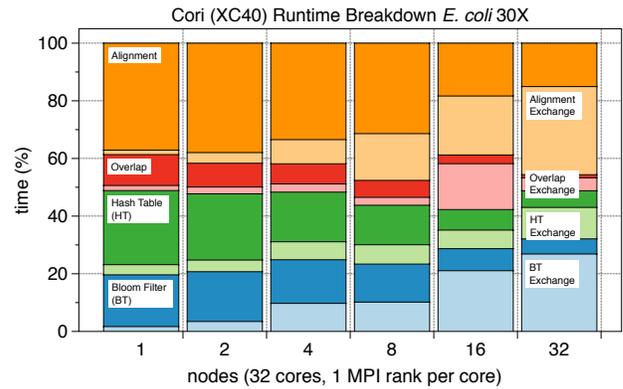}
  \caption{Runtime breakdown on Cray XC40 with minimum computational-intensity workload (E.coli 30x, single seed).}%
  \label{fig:coril_pct_runtime}
\end{figure}

\begin{figure}
  \centering 
  \includegraphics[width=\columnwidth]{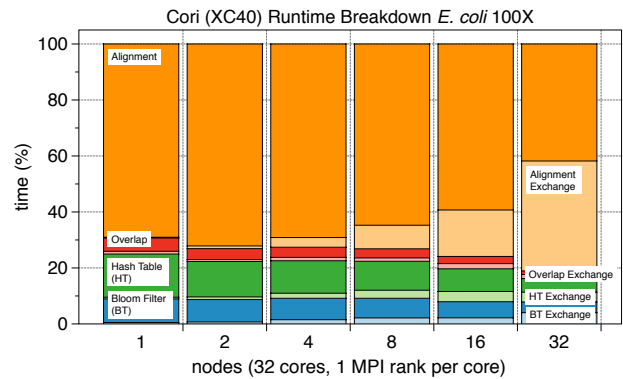}
  \caption{Runtime breakdown on Cray XC40 with higher computational-intensity workload (E.coli 100x, all seeds separated by at least 1Kbps).}%
  \label{fig:corih_pct_runtime}
\end{figure}

\begin{figure}
  \centering 
  \includegraphics[width=\linewidth]{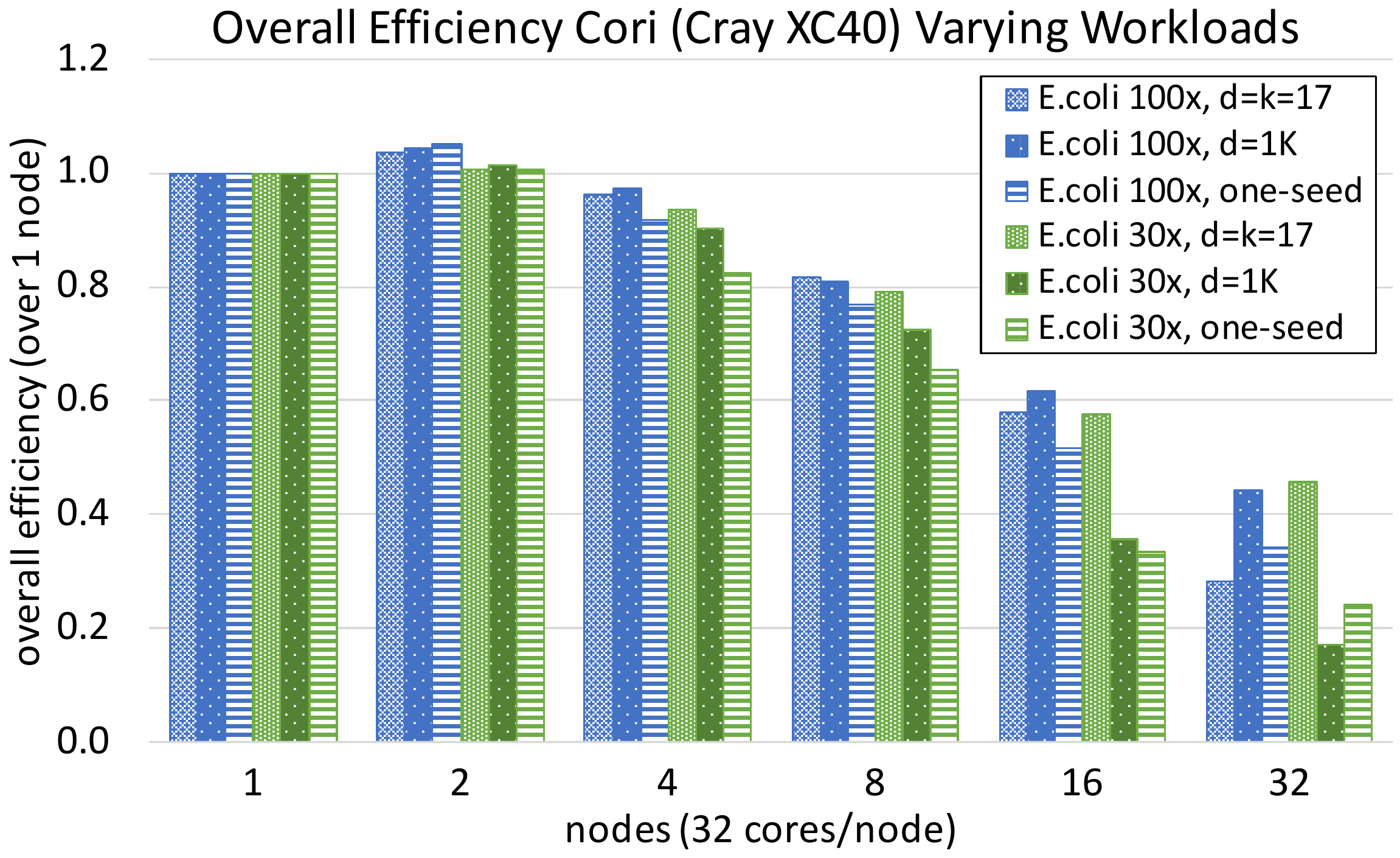}
   \caption{\Marquita{Overall efficiency on Cray XC40 over 2 data sets, E.coli 30x and E.coli 100x, varying seed constraints (1 seed, all separated by 1K, and all separated by k=17 characters).}}%
  \label{fig:cori_varying_workload}
\end{figure}

The performance rates on each stage show similar results across machines, with the more powerful Haswell CPU nodes and network on Cori (XC40) giving superior overall performance.  As expected, all-to-all style communication scales poorly on all networks, with but especially the commodity AWS network.  Somewhat more surprising is the high level of superlinear speedup on some stages once the data fits in cache or other memory hierarchy level. The question for overall performance is how these two effects trade off against one another and how the stages balance out. 
\Marquita{Figure~\ref{fig:cori_varying_workload} shows diBELLA's overall pipeline efficiency on Cori, varying workloads and computational intensity. Two data sets are used, \ecoli 30x and \ecoli 100x, and 3 seed constraints, one-seed, all seeds separated by 1Kbps, and all seeds separated by $k=17$ bps. Clearly, increasing the computational intensity with larger inputs and seed counts does not alone determine overall efficiency. While the computational efficiency increases with higher computational intensity, the overall efficiency is significantly impacted by the degrading efficiency of exchanges.}
Figures~\ref{fig:coril_pct_runtime} and~\ref{fig:corih_pct_runtime} respectively show runtime breakdown by stages on Cori for \ecoli $30\times$ exploring 1 seed per overlapping pair of reads, and \ecoli $100\times$, using all seeds with a minimum of 1Kbps-distant from each other. These represent two extremes in terms of computational intensity, the former being minimal, the latter being much higher, but still a realistic point of comparison for the same input genome.
The communication time is broken out for each stage. The stages are fairly evenly balanced, although alignment is more expensive computationally than the others (and dominates to 32 nodes in the more computationally intense workload). Focusing on Figure~\ref{fig:coril_pct_runtime}, the communication time in the Bloom Filter stage is surprisingly higher than in the Hash Table stage where the volume is 2.5x higher, and the communication pattern and number of messages is identical.  Further investigation revealed that the problem is the first call to the MPI Alltoallv routine, which is almost twice as expensive the first time as the second, so the Hash Table stage benefitted from whatever internal data structure and communication initialization happened in the Bloom Filter stage. This effect was visible to varying degrees on all 4 platforms. This kind of behavior is most noticeable for workloads with lowest computational intensity.

\begin{figure}
  \centering 
  \includegraphics[width=\linewidth]{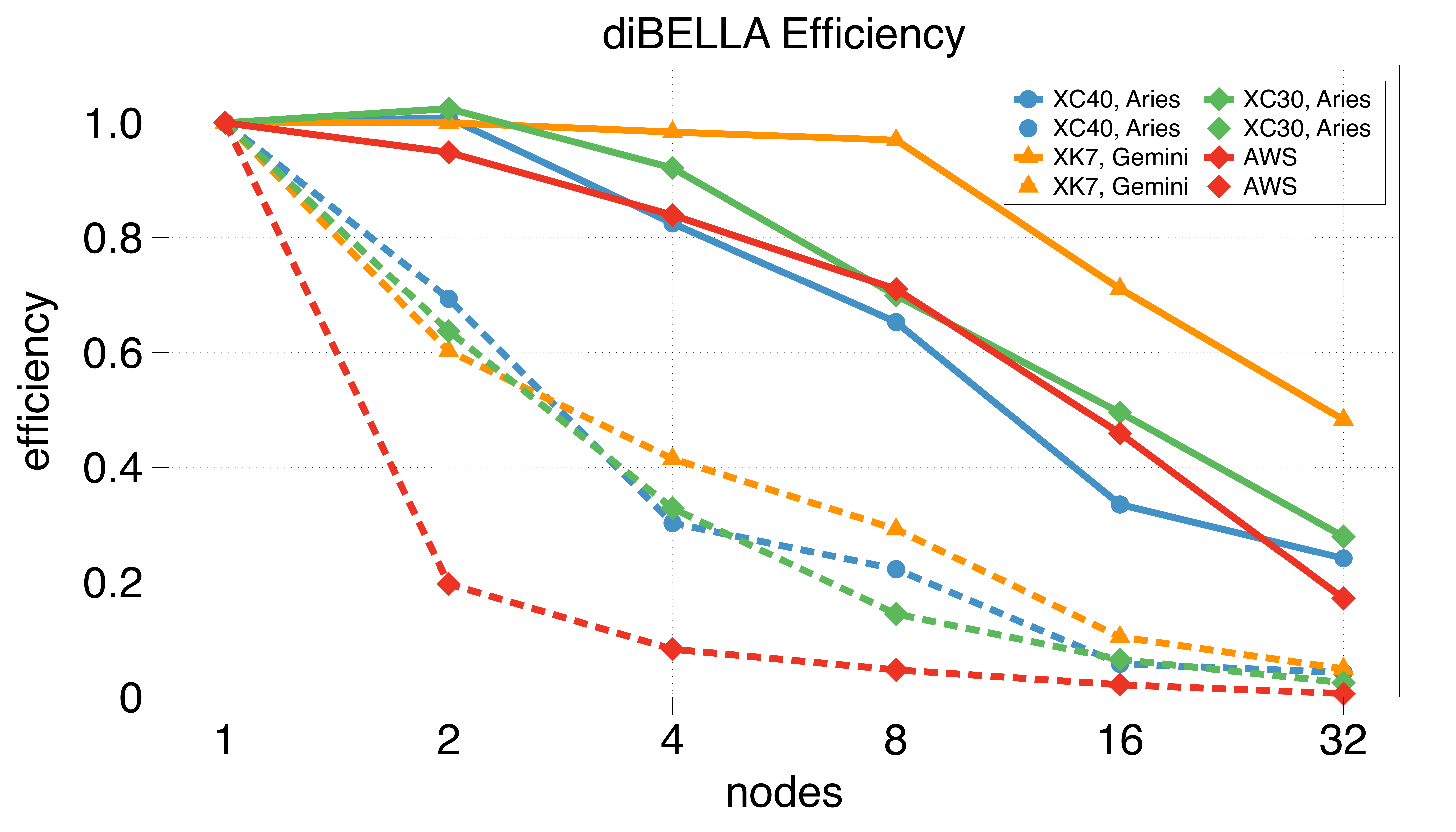}
  \caption{DiBELLA cross-architecture overall (solid) and exchange (dashed) efficiency over minimally compute intensive workload (E.coli 30x one-seed).}%
  \label{fig:cross_hpc_network}
\end{figure}

To further drill down on network and processor balance, 
Figure~\ref{fig:cross_hpc_network} shows the efficiency across all 3 HPC networks over the overall pipeline and the exchange time on each.  From an efficiency standpoint, the Cray XK7 using only the CPU features on each node gives the best network balance for this problem, even though the network is an older generation than on the XC30 and XC40.

\begin{figure}
  \centering 
  \includegraphics[width=\linewidth]{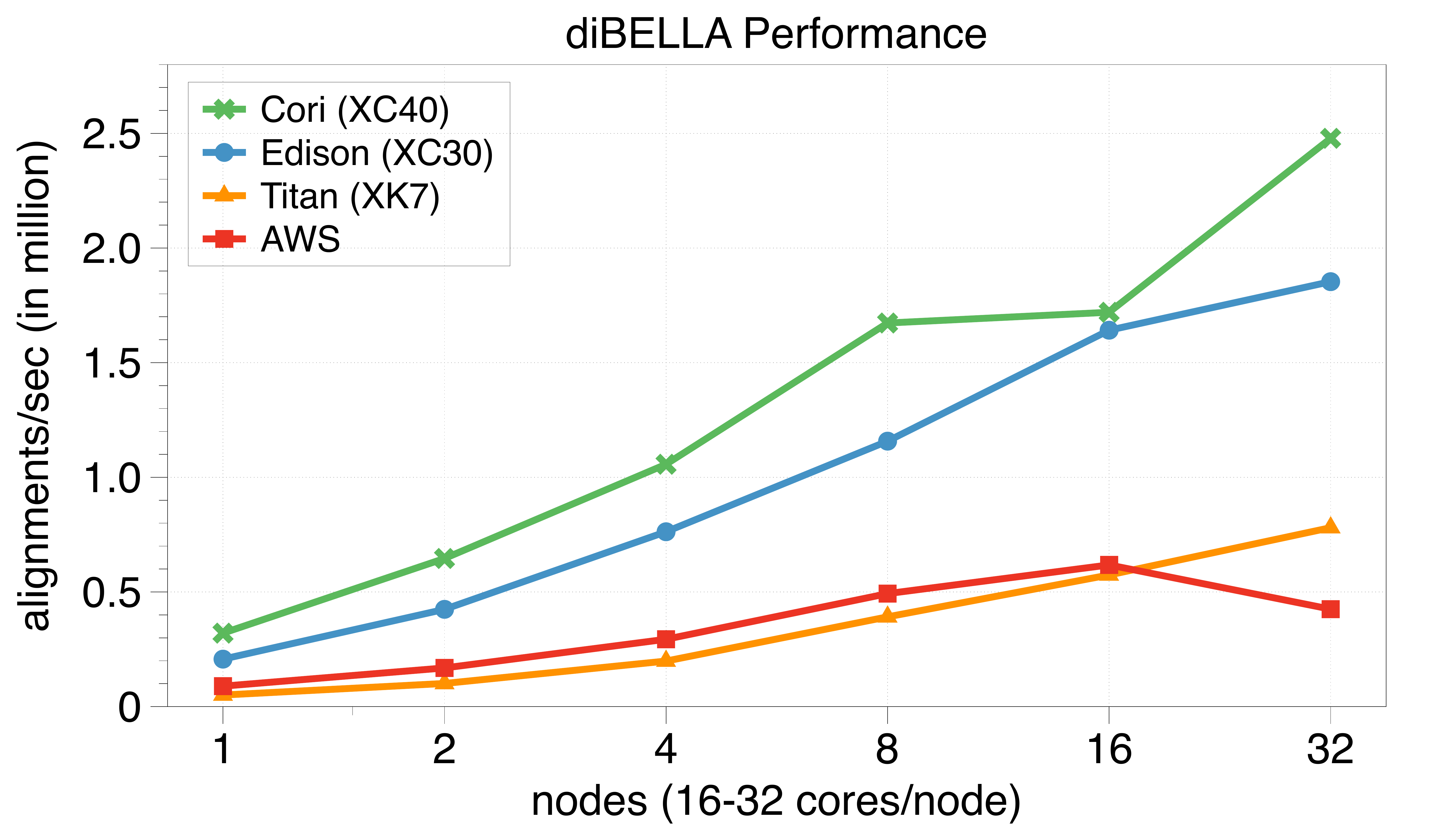}
   \caption{DiBELLA cross-architecture strong scaling as millions of alignments per second given E.coli 30x one-seed.}%
  \label{fig:pipeline_rate}
\end{figure}

From a performance standpoint, the higher speed processor and network on Cori (XC40), while not as well balanced for efficiency, outperforms the other on the full diBELLA pipeline, Figure~\ref{fig:pipeline_rate}. Here we measure performance as alignments per second, where the total number of alignments is fixed for a given input configuration.  The performance anomaly at 16 nodes for Cori, which was seen in the Alignment stage, is apparent here in the overall performance, probably due to a performance issue in MPI implementation.  With that exception and a drop of performance on AWS at 32 nodes, all of the systems show increasing performance on increased node counts.  Recall, that our standard problem used here (\ecoli at 30x coverage with only 1 seed used per read pair for alignment) was specifically chosen as the low end of computational intensity, and so highlights scaling limits of the machines.

\section{Related Work}\label{sec:prior}
Nowadays, existing distributed memory alignment codes target the alignment of a read set against a fixed, modest-sized reference sequence such as the human genome~\cite{guo2018bioinformatics}, where the reference can be replicated across nodes in advance.
Conversely, diBELLA computes read-to-read pairwise alignment rather than read-to-reference alignment, and distributes data across nodes for each step of the pipeline.
Our implementation is more similar (in spirit) to the end-to-end parallelization in MerAligner~\cite{georganas2015meraligner}. 
\Marquita{ However, diBELLA addresses long read data characteristics, and accordingly, uses a different parallelization and data layout approach.
MerAligner aligns short reads to {\it contigs}, sequences composed of error-free \kmers, in order to find connections between {\it contigs}. Long reads are not only 2-3 orders of magnitude longer than short reads, but also contain errors up to 35\% 
(versus < 1\% for short reads). Hence, appropriate \kmer lengths for long read overlap and alignment are an order of magnitude shorter. These features combined dramatically increase the size of the \kmer data set.  
Further, in MerAligner, the cardinality of the {\it contig} set is reduced significantly over the size of the input (see \cite{egeorganas2016} and \cite{mme2017}); whereas in long-read-to-long-read alignment, a potentially all-to-all comparison of input reads may be performed.  
In summary, these differences result, for the long read case, in significantly 
higher communication, computation, and memory consumption rates, and a fundamental difference in pairwise alignment kernel and parameter choices (relating both to quality and computational cost). While exploring a PGAS design for this application similar to MerAligner's is part of ongoing work, we do not provide a direct comparison since they target significantly different problems.
}

\begin{table}[t]
\centering
\caption{ Single node, 64 thread runtime (s) comparison (excluding I/O) on Cori Haswell w/ 128 GB RAM. Reported DALIGNER times also exclude all pre- and post- processing.}%
\label{tab:runtime_1node}
\begin{adjustbox}{width=\columnwidth}
\begin{tabular}{llll}
\hline\noalign{\smallskip}
 & {\it E.coli} 30x (sample) & {\it E.coli} 30x & {\it E.coli} 100x \\
\hline\noalign{\smallskip} 
diBELLA & 12.74 & 65.72 & 79.45 \\
DALIGNER & 7.31 & 52.04 & 63.70 \\
\hline
\end{tabular}
\end{adjustbox}
\end{table}

In addition to BELLA, which is the method that our distributed algorithm is based on, other state-of-the-art long-read to long-read aligners include BLASR~\cite{blasr}, MHAP~\cite{berlin2015assembling}, Minimap2~\cite{li2018minimap2}, MECAT~\cite{xiao2017mecat}, and DALIGNER~\cite{myers2014efficient}.
BLASR uses BWT and FM index to identify short \kmer matches and then groups \kmer matches within a given inter-\kmer distance.
Grouped \kmers are ranked based on a function of the \kmer frequency and highly scored groups are kept for downstream analysis.
MHAP, Minimap2, MECAT, and DALIGNER use \kmer matches for identifying candidate overlapping pairs, similarly to diBELLA.
MHAP computes an approximate overlap detection performing sketching on the \kmer set using minHash.
Compact sketches are used to estimate the similarity between sequences.
Minimap2 uses minimizers rather than all possible \kmers to obtain a compact representation of the original read.
Collinear \kmers on a read are chained together and used for finding possible matches with other sequences.

Like BELLA and diBELLA, MECAT and DALIGNER do not use approximate representations. 
MECAT divides reads into blocks and scans for identical \kmers which are used to calculate a distance difference factor first between neighboring \kmers hits, and then between neighboring blocks.
DALIGNER computes a \kmer sorting based on the position within a sequence and then uses a merge-sort to detect common \kmers between sequences.
\Marquita{DALIGNER supports problem sizes exceeding single node resources through a scripting frontend, that divides work into a series of independent execution steps. This script-generated-script can be executed directly (serially) on a single node, or the user can modify it to run independent batch jobs, as individual node resources become available in a distributed setting.}
For example, if the data is divided into blocks, $B_1,...B_n$ then DALIGNER can be executed independently by aligning $B_1$ to itself as one job, $B_2$ to itself and $B_1$ as another, and so on.
This approach addresses the memory limitations, but it is not scalable.
DALIGNER's distributed memory approach reads $B_1$ from disk $n$ times and the amount of work varies significantly across nodes. \Marquita{Given these differences, we do not provide a direct multi-node comparison, however for completeness, we provide a single node runtime comparison of diBELLA and DALIGNER in Table~\ref{tab:runtime_1node}. We exclude I/O time from each measurement, since each tool handles both input and output in significantly different ways. For DALIGNER, we additionally exclude all preprocessing (initializing the database and splitting it into blocks) and post processing (all commands for verifying and merging results) time, and implicitly, the time required of DALIGNER users to extract human-readable results of interest from the database. Table~\ref{tab:runtime_1node} shows that diBELLA's single node runtime is competitive with DALIGNER's across these data sets, even excluding DALIGNER's I/O, preprocessing, and postprocessing.}

\Marquita{
Of the alternatives, BELLA is the latest (and ongoing) work, with a comprehensive quality comparison to all the above \cite{guidi2018bella}. BELLA's quality is competitive, especially excelling in comparisons where the ``ground truth" is known. Further, BELLA offers a computationally efficient approach, yielding {\it consistently} high accuracy across data sets, and a well-explained and supported methodology. For these reasons, we chose BELLA as the basis for our distributed memory algorithm. The quality produced by diBELLA is at least that of BELLA (see \cite{guidi2018bella} for quality comparisons over data sets also used in this study), and higher when using less restricted sets of seeds than \cite{guidi2018bella}.  }

{\it De novo} genome assembly depending on long-read alignment is becoming a crucial step in bioinformatics. 
Current available long-read {\it de novo} assemblers are Canu~\cite{koren2017canu}, which uses MHAP as long-read aligner, Miniasm~\cite{li2016minimap} which uses Minimap2, and HINGE~\cite{kamath2017hinge} and FALCON~\cite{chin2016phased}, which use DALIGNER.
Flye~\cite{kolmogorov2019assembly} uses the longest jump-subpath approach~\cite{lin2016assembly} to compute alignments. 
From a hardware acceleration standpoint, there has been increasing interest in the long read alignment problem as well, as evidenced by recent work \cite{turakhia2018darwin}\cite{alser2019shouji}. Though we do not explore it in this work, we leave it as a promising future direction.

As mentioned in Section~\ref{sec:background}, long-read to long-read alignment requires filtering out part of the \kmers in order to avoid either spurious alignments or performing unnecessary computation.
The parallel \kmers analysis in diBELLA is built upon that of HipMer~\cite{georganas2018hipmer}.
The Bloom filter stage is identical, while the hash table implementation is different.
For each \kmer, HipMer stores only the two neighboring bases in the original read it was extracted from (which have to be unique, so there is a single entry per \kmer).
diBELLA instead needs to communicate and store information about the read where the \kmer originated and the location in which each \kmer $\ ${\it instance} appeared.
Both HipMer and diBELLA remove singleton \kmers, but diBELLA also removes those \kmers whose occurrence exceeds the high occurrence threshold, $m$.
The hash tables also represent different objects.
The HipMer hash table represents a de Bruijn graph with \kmer vertices, and their connections are computed by adding the \kmer extensions and shifting.
The graph is broken at points where there is no confidence in the most likely extension. The high error rates in long reads would make such a graph very fragmented.
DiBELLA's hash table represents a read graph with read vertices connected to each other by shared \kmers, the \kmers are then extended to overlapping regions through pairwise alignment.
This graph representation, often known as the {\em overlap graph} in the literature,  
is more robust to sequencing errors and thus more suitable for long-read data.
T. Pan et al's work \cite{pan2018optimizing} provides interesting optimizations to HipMer and others' distributed memory \kmers analysis design, but focuses on the general problem of counting \kmers in distributed memory, and not also the construction and computation on the read overlap graph.


\section{Conclusions}\label{sec:conclusions}

We presented diBELLA, a long read to long read aligner for distributed memory platforms that deals with the unique problem of aligning noisy reads to each other, making it possible to analyze data sets that are too large for a single shared memory and or making heroic computations routine.   Alignment is a key step in long read assembly and other analysis problems, and often the dominant computation. 
diBELLA avoids the expensive all-to-all alignment by looking for short, error-free seeds (\kmers) and using those to identify potentially overlapping reads.  We believe this is the first implementation of such a long read-to-read aligner designed for distributed memory.  In addition to the independent work of performing pairwise alignment on many reads, our implementation takes advantage of global information across the input data set, such as the total count of each \kmer used for filtering errors and the distribution of overlaps to distribute load.  We performed a thorough performance analysis on 3 leading HPC platforms as well as one commodity cloud offering, showing good parallel performance of our approach, especially for realistic scenarios that perform multiple alignments per pair of input reads.  While the HPC systems offer superior performance to the cloud, all of them benefit from the multi-node parallelization.  The application is dominated by irregular all-to-all style communication and the study reveals some of the performance anomalies on particular systems, as well as general scaling issues at larger node counts.

We believe that in addition to being a useful tool for bioinformatics, either standalone or as part of a larger pipeline, diBELLA also represents an important parallel workload to understand and drive the design of future HPC system and communication libraries, and a basis for future optimizations in single node optimization for pairwise alignment while retaining the efficiency sparse nature of the interactions.

\begin{acks}
	This work was supported in part 	
         by the Exascale Computing Project
            (17-SC-20-SC), a collaborative effort of the U.S. Department
           of Energy Office of Science and the National Nuclear Security
           Administration
          and by the National Science Foundation.  
    This research used resources of the National Energy Research Scientific
         Computing Center (NERSC) under contract No. DE-AC02- 05CH11231,
    and the Oak Ridge Leadership Computing Facility under DE-AC05-00OR22725.
    all supported by the Office of Science of the U.S. Department of Energy.
        The information presented here does not necessarily 	
	reflect the position or the policy of the Government 	
	and no official	endorsement should be inferred. 
       AWS Cloud Credits were provided through Amazon Web Services and PI
       Benjamin Brock.   

\end{acks}

\bibliographystyle{ACM-Reference-Format}
\bibliography{../refs}
\end{document}